\documentclass{ifacconf}

\usepackage{graphicx}
\usepackage{natbib}
\usepackage{amsmath,amssymb}
\usepackage{booktabs}
\usepackage{xcolor}

\newcommand{\R}{\mathbb{R}}

\begin{document}
\begin{frontmatter}

\title{Multi-experiment identification from short repeated transients:
the human response to pushes in a Tai Chi posture}

\author[First]{Emmanuel Witrant}
\author[First]{Olivier Martin}
\author[First]{Pascal Bellemain}

\address[First]{Univ. Grenoble Alpes, CNRS, GIPSA-lab,
38000 Grenoble, France (e-mail: emmanuel.witrant@univ-grenoble-alpes.fr,
ORCID 0000-0002-2026-8915; Olivier.Martin@gipsa-lab.fr;
Pascal.Bellemain@gipsa-lab.fr).}

\begin{abstract}
The effect of Tai Chi postures on the mechanical response of the body to
external pushes is usually assessed with clinical balance scores, which
say little about the underlying dynamics. We identify, for five subjects
and eight postural conditions, linear models from the mechanical work
transferred by a pendulum hitting the forearm to the change of the
antero-posterior ground reaction force. Each trial contains 10 to 16
impacts, and each impact yields a transient of 0.55\,s only. Building on
earlier analyses of these data, we improve the processing chain in three
ways. The input is related to the energy lost by the pendulum and its
repeatability is quantified. The output is defined component-wise, which
makes it independent of the zeroing of the force plate. Finally, the
impacts are stacked in a multi-experiment least-squares problem and
validated on unseen impacts. Tenth-order ARX models then reach a median
leave-one-impact-out fit of 65\%, against 72\% for the repeatability of
the response itself, and never require stability enforcement, unlike
estimation on averaged impacts. Closing the eyes lengthens the
identified delay by about 70\,ms in all subjects, and the Tai Chi
posture increases the antero-posterior force transmitted per joule in
four of five subjects. Surface EMG envelopes show a tonic co-activation
of the lower-leg muscles in this posture, a smaller reactive triceps
burst and a fast habituation of the tibialis response. Together, they
point to a stiffer, anticipatory postural strategy.
\end{abstract}


\end{frontmatter}

\section{Introduction}

Tai Chi Chuan is widely practised as a low-impact exercise, and several
meta-analyses report that it reduces the risk of falls in older adults
\citep{LomasVega2017,Huang2017}. The evidence is nevertheless contested:
overviews of systematic reviews point to heterogeneous protocols, small
samples and possible publication bias \citep{LeeErnst2012}, and the
balance parameters used as outcomes vary widely between studies
\citep{JimenezMartin2013}. Most of these outcomes are clinical scores or
static posturographic indices, which quantify \emph{whether} balance is
maintained, not \emph{how} the body dynamically transmits and rejects a
perturbation.

Answering the latter question requires an input--output model of the
perturbed body. Musculoskeletal and multibody models
\citep{Delp2007,Ghafari2007,Rajagopal2016} are physically detailed but
involve far more parameters than a few external measurements can
identify. Postural-control models based on inverted pendulums, identified
from continuous support-surface or visual perturbations
\citep{Peterka2002,vanderKooij2005,Kiemel2008}, are parsimonious but
target quiet stance with the perturbation applied at the feet. Neither
addresses a push applied on the \emph{upper} limb of a subject in a
flexed stance such as the Tai Chi ``Peng'' posture, where the
perturbation travels through the arm, trunk and legs before reaching the
ground.

We therefore follow a black-box route, which raises identification
issues typical of biomechanical experiments: the excitation is a short,
nearly repeatable transient, and each trial provides only a dozen of
them. The responses vary from one impact to the next, and the number of
subjects is small. The paper makes four contributions. The first is a
complete processing chain from raw motion-capture and force-plate data to
input--output transients. It uses an energy-based input and an impact
detection based on the sign change of the pendulum power, quantifies the
repeatability of the input, and defines the force output
component-wise. The second is a comparison between identification on
the ensemble-averaged transient and multi-experiment identification on
individual impacts. Both are assessed by leave-one-impact-out
cross-validation, by information criteria, by a repeatability bound
computed from the data alone, and by their sensitivity to the length of
the analysis window. The third is a set of model-based descriptors
(delay, force gain, time to peak) that remain interpretable for
high-order models. They are compared across eight postural conditions,
five subjects and two levels of expertise. The fourth is an analysis of
surface EMG envelopes of arm and leg muscles, both as descriptors of the
neuromuscular strategy and as additional inputs of multi-input models.
The experiments were first analysed in a Master's thesis \citep{NYo2019}
and in a short conference contribution \citep{Martin2023}, which showed,
for the Tai Chi conditions and one expert, that tenth-order ARX models
fitted on averaged impacts reproduce the GRF response to the pendulum
work and its modulation by load and vision. Transfer-function analyses
of balance recovery had also been applied to vestibular patients
\citep{Voda2019}, and multi-joint identification of standing balance to
continuous perturbations \citep{Engelhart2015}. The present paper
extends \citet{Martin2023} to all subjects and conditions. It refines
the definition of the input and output signals and the estimation
strategy, validates the models on unseen impacts, and adds
interpretable descriptors, the muscle activity and the adaptation across
impacts.

Section~\ref{sec:exp} describes the experiment,
Section~\ref{sec:signals} the construction of the signals and the
repeatability of the input, Section~\ref{sec:ident} the identification
and validation methods and Section~\ref{sec:results} the results, which
are discussed in Section~\ref{sec:discussion}. Code and data are openly
available \citep{Witrant2026data}.

\section{Experimental protocol}\label{sec:exp}

Five adult volunteers (one woman, four men; body mass 73--105\,kg,
height 1.68--1.90\,m, age 30--58 years where recorded) took part. Two of them
(S1, S2) are experienced Tai Chi practitioners. All participants gave
their informed consent according to ethical research standards
\citep{Martin2023}.

The experiments were carried out on the biomechanics platform of
GIPSA-lab. The subject stands on a force plate (Kistler 9286A) that
measures the ground reaction force (GRF)
$F(k)=[F_x(k),F_y(k),F_z(k)]^\top$. The width of the stance is therefore
limited by the size of the plate. The subject holds a defence attitude, with the
right arm--hand segment placed forward at shoulder height to resist the
impact. A pendulum is released from a fixed (but not recorded) height
towards this segment and is released by a manually activated
quick-release device after the impact \citep{Martin2023}. Throughout
the trials, the load is applied at the forearm--hand level. Five
reflective markers fixed on the pendulum are tracked by eight infrared cameras
(Qualisys Oqus 600). Force-plate and marker data are recorded
synchronously at $f_s=100$\,Hz ($T_s=0.01$\,s) over trials of 40 or
45\,s. As set in the acquisition software, the force plate is zeroed at
the beginning of each recording, with the subject standing on it.
Surface EMG of four muscles (right triceps brachii and biceps
brachii, left tibialis anterior and gastrocnemius) was recorded with a
Biopac MP150 system (EMG100C amplifiers) through the analog inputs of the
motion-capture system, at the same 100\,Hz rate. The 33 body markers
are not used here.

\begin{figure}
  \centering
  \includegraphics[width=0.75\columnwidth]{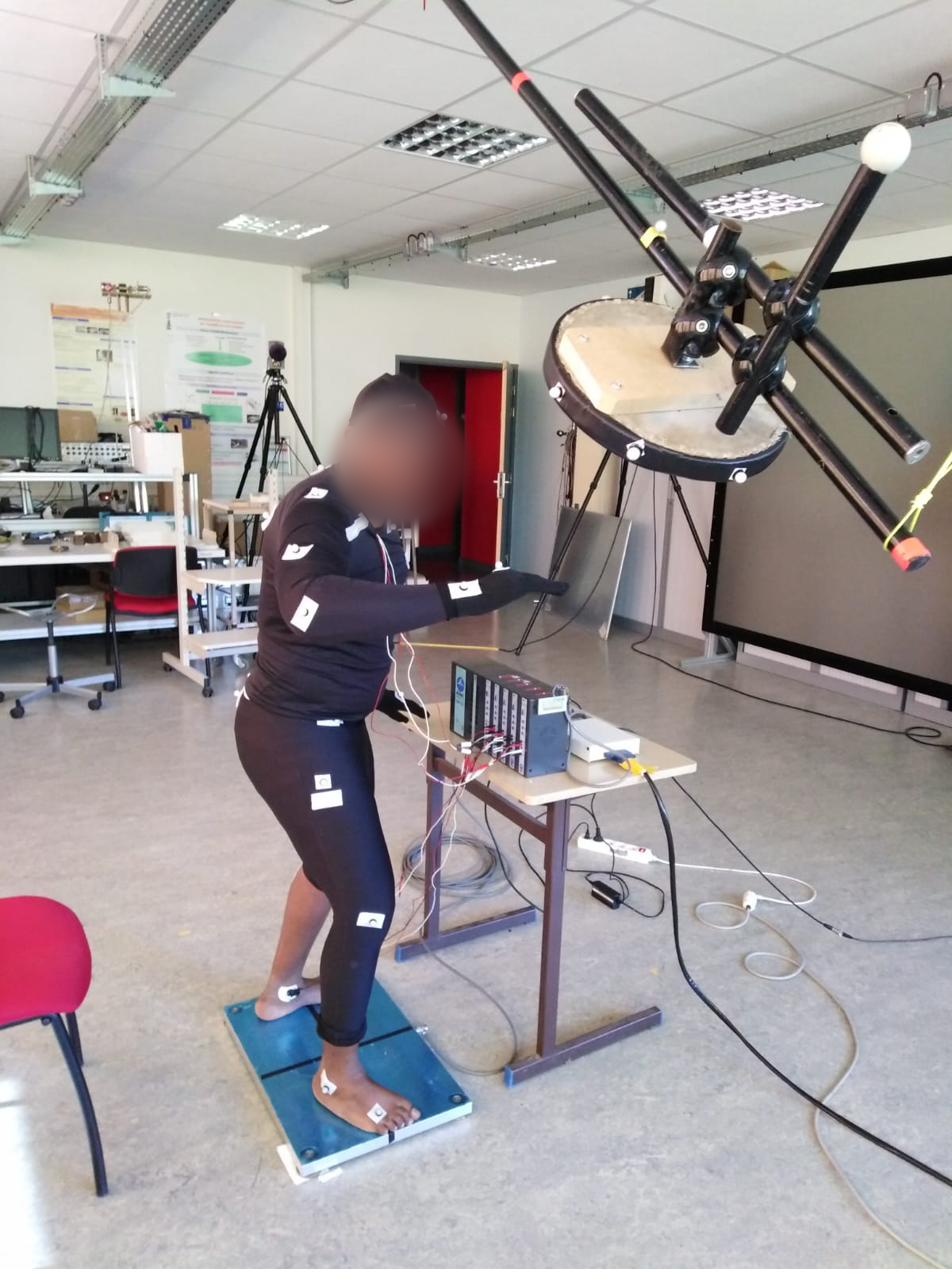}
  \caption{Tai Chi ``Peng'' posture adopted in the TS conditions.}
  \label{fig:setup}
\end{figure}

Each subject was recorded in the eight conditions of
Table~\ref{tab:conditions}, with a few minutes of rest between trials:
normal stance with feet apart (NS) or together (CF), Tai Chi posture (TS, Fig.~\ref{fig:setup}),
eyes open (OE) or closed (CE), and a heavier pendulum (WL). The Tai Chi posture with
eyes open (TSOENL) is the reference condition. The pendulum mass is
$m_p=4.5$\,kg. In the two loaded conditions, a 2-kg load is added to the
pendulum ($m_p=6.5$\,kg), so that the subject receives a larger impact
energy in the same posture. The conditions were recorded in the order of
Table~\ref{tab:conditions}.

\begin{table}
\caption{Experimental conditions ($^{*}$reference).}\label{tab:conditions}
\centering\footnotesize\setlength{\tabcolsep}{4pt}
\begin{tabular}{lll}
\toprule
Label & Stance & Eyes, pendulum \\
\midrule
NSOENL  & normal, feet apart     & open, 4.5\,kg \\
NSCENL  & normal, feet apart     & closed, 4.5\,kg \\
CFOENL  & normal, feet together  & open, 4.5\,kg \\
CFCENL  & normal, feet together  & closed, 4.5\,kg \\
TSOENL$^{*}$ & Tai Chi (Peng)    & open, 4.5\,kg \\
TSCENL  & Tai Chi (Peng)         & closed, 4.5\,kg \\
TSOEWL  & Tai Chi (Peng)         & open, 6.5\,kg \\
TSCEWL  & Tai Chi (Peng)         & closed, 6.5\,kg \\
\bottomrule
\end{tabular}
\end{table}

\section{Construction of the input--output transients}\label{sec:signals}

\begin{figure}
  \centering
  \includegraphics[width=\columnwidth]{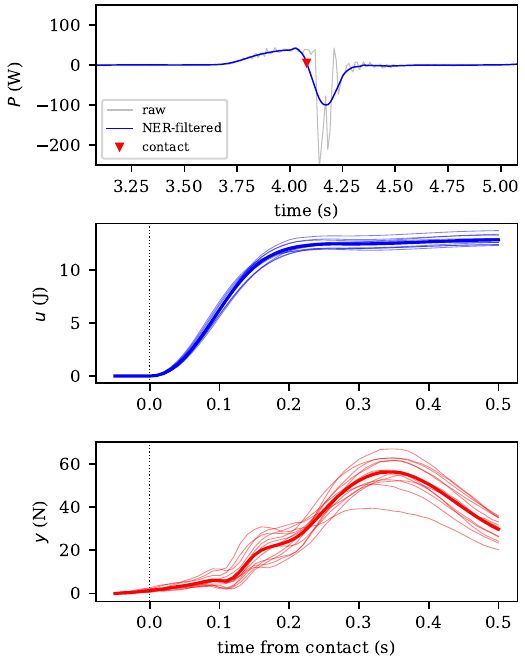}
  \caption{Signal construction (S1, TSOENL). Top: raw and NER-filtered
  pendulum power around one impact, with the detected contact instant.
  Middle: transferred work $u_i$ for the 13 impacts (thin) and their mean
  (thick). Bottom: change of the antero-posterior GRF $y_i$.}
  \label{fig:pipeline}
\end{figure}

The input is computed from the motion of the pendulum. Let $p(k)\in\R^3$ be the position (m) of the
first pendulum marker that is valid over the whole trial (a marker is
discarded if its trajectory contains two consecutive zero samples). Its
acceleration is obtained by central second differences, with one-sided
four-point formulas at the ends, and its velocity by backward
differences,
\begin{align}
  a(k)&=\frac{p(k+1)-2p(k)+p(k-1)}{T_s^2},\\
  v(k)&=\frac{p(k)-p(k-1)}{T_s}.
\end{align}
The instantaneous power of the pendulum is
\begin{equation}
  P(k) = m_p\, a(k)^\top v(k)\quad[\mathrm{W}],
\end{equation}
positive while the pendulum gains kinetic energy and negative when it is
decelerated by the arm. The double differentiation amplifies the marker
noise (Fig.~\ref{fig:pipeline}, top), so $P$ is smoothed by the
``nearly equal ripple'' (NER) low-pass filter of \citet{KaiserReed1977},
a symmetric, zero-phase FIR filter designed with a Kaiser window. With
a pass-band edge $\beta=0.1$ (normalised to the Nyquist frequency, i.e.\
5\,Hz), a transition width $\delta=0.1$ and a ripple $\varepsilon=0.06$,
the design gives $n_p=12$ coefficient pairs (25 taps). The $n_p$ samples
at each end of the record are replaced by centred moving averages over
at most 17 samples.

Impacts are first detected when
$P$ exceeds 30\,W, starting 0.5\,s after the beginning of the trial and
stopping 1.5\,s before its end, with a refractory period of 2\,s. Each
detection $\tilde k_i$ is then refined as the instant of maximal
cumulative power over the interval $[\tilde k_i-15,\tilde k_i+35]$,
\begin{equation}
  k_i=\arg\max_{k}\;\sum_{j=\tilde k_i-15}^{k} P(j),
\end{equation}
i.e.\ the instant where $P$ changes sign and the pendulum kinetic energy
is maximal, which we take as the contact instant. Each impact provides a
window of $n_t=56$ samples, from $k_i-5$ to $k_i+50$ (0.05\,s before
to 0.5\,s after).

The input for the identified model is the work transferred from the
pendulum after contact,
\begin{equation}
  u_i(t)= -T_s\sum_{j=k_i+1}^{k_i+t} P(j),\qquad
  t=1,\dots,50,
  \label{eq:work}
\end{equation}
and $u_i(t)=0$ before contact. Integrating the power turns the brief,
noisy power pulse into a smooth, step-like signal (Fig.~\ref{fig:pipeline},
middle). For the output, the zeroing of the force plate matters: at
rest, $F$ is not the body weight but a small residual (5 to 45\,N,
mostly vertical, of either sign) that drifts during the trial. The
earlier analyses used the change of the norm,
$\|F(t)\|_2-\|F(k_i-5)\|_2$. With a zeroed plate, a transient decrease
of $F_z$ produces a dip of this norm when the residual is positive and a
peak when it is negative, so that the output depends on the zeroing
offset. We therefore use the change of the
antero-posterior component, in the direction of the push,
\begin{equation}
  y_i(t)=F_y(k_i+t)-F_y(k_i-5)\quad[\mathrm{N}],
\end{equation}
which rises smoothly after contact in all trials
(Fig.~\ref{fig:pipeline}, bottom). The vertical component, which shows a
faster biphasic transient, and the small medio-lateral one are not
modelled here. For each subject and condition, the $N_I$ impacts form the data set $\{(u_i,y_i)\}_{i=1}^{N_I}$.

Surface EMG has most of its power between 20 and
450\,Hz and is normally sampled at 1--2\,kHz. At 100\,Hz, the recorded
signals are aliased into the 0--50\,Hz band, and several channels are
dominated by mains interference folded close to the Nyquist frequency
(up to 98\% of the power above 48\,Hz). Because aliasing without
anti-aliasing filter preserves the signal power, the amplitude envelope
still reflects muscle activation, but the waveform and the fine timing
of the bursts are lost. Each channel is therefore band-pass filtered
between 10 and 40\,Hz (fourth-order Butterworth, zero phase), which
removes motion artefacts and the folded mains, full-wave rectified and
low-pass filtered at 6\,Hz (second order, zero phase). In the absence
of maximal voluntary contractions, two normalisations are used: the
\emph{phasic} response of an impact is the envelope divided by its mean
over the 0.45\,s preceding the window, and the \emph{tonic} activity of
a trial is the median envelope outside $[k_i-0.2\,\mathrm{s},
k_i+0.8\,\mathrm{s}]$, compared between conditions of the same subject
(same electrodes) relative to NSOENL.

We now assess the repeatability of the input. Within a trial, subject
and posture are fixed and the pendulum is released from the same height,
so the inputs should be identical. In practice they are not. To
understand why, note that with the differentiation scheme above
$a(k)=(v(k+1)-v(k))/T_s$, so that, before smoothing, \eqref{eq:work}
satisfies exactly
\begin{equation}
  u_i(t) = \Delta E_i(t)
  + \frac{m_p}{2}\sum_{j=k_i+1}^{k_i+t}\|v(j+1)-v(j)\|^2 ,
  \label{eq:decomp}
\end{equation}
where $\Delta E_i(t)=E(k_i+1)-E(k_i+t+1)$ is the loss of kinetic
energy $E=\frac{m_p}{2}\|v\|^2$ of the pendulum. The second term is a
discretisation artefact: it is always positive, grows with the roughness
of the velocity, and therefore amplifies marker noise. Within the
0.5\,s window, the pendulum loses almost all its kinetic energy: over
the 482 impacts, the ratio between the energy lost and the energy at
contact $E(k_i)$ has a median of 0.997 and exceeds 0.98 for 95\% of the
impacts. The kinetic energy at contact is also very repeatable within a trial (median coefficient of variation 4\%),
which confirms that the release was well controlled. The work $u_i$
exceeds $\Delta E_i$ by a median of 45\%, and its within-trial
coefficient of variation is twice as large (median 8\%, up to 72\%). In
33 impacts, the artefact more than doubles the input, when the velocity
is rough after contact (e.g.\ S1, CFCENL).

We finally examine whether the amplitude of the transmitted energy
depends on the subjects and postures. The energy at contact ranges from 7 to 16\,J between trials
(9.3\,$\pm$\,1.9\,J with the light pendulum, 13.0\,$\pm$\,1.3\,J with
the heavy one). Since the release height is fixed, the variations
between trials of the same mass come from the point of the swing at
which the forearm intercepts the pendulum, which depends on the subject
and on the posture. The input is thus not strictly exogenous. A
two-way analysis of variance on the individual impacts, with subject and
condition as factors, tests these effects on the energy at contact, the
peak absorbed power and the absorption duration. The $p$-values of the
$F$-tests of the two main effects are below $10^{-15}$ in all cases,
i.e.\ both effects are significant. The interaction term, however, is of
the same order as the main effects (25\% to 45\% of the total
variance). Posture
effects on the transmitted energy are therefore subject-specific, and a
general conclusion would require a linear mixed-effects model
\citep{PinheiroBates2000}, with posture, vision and load as fixed effects
and the subject as a random effect, fitted on a larger cohort.

\section{Identification and validation}\label{sec:ident}

The dynamics from $u$ to $y$ is described by an
ARX model \citep{Ljung1999}
\begin{equation}
  A(q^{-1})\,y(t) = q^{-n_k} B(q^{-1})\,u(t) + e(t),
  \label{eq:arx}
\end{equation}
with $A(q^{-1})=1+a_1q^{-1}+\dots+a_{n_a}q^{-n_a}$,
$B(q^{-1})=b_1+\dots+b_{n_b}q^{-n_b+1}$ and $e$ a white noise sequence.
With $\theta=[a_1,\dots,a_{n_a},b_1,\dots,b_{n_b}]^\top$ and the
regressor $\varphi(t)=[-y(t-1),\dots,-y(t-n_a),u(t-n_k),\dots,
u(t-n_k-n_b+1)]^\top$, \eqref{eq:arx} reads
$y(t)=\varphi(t)^\top\theta+e(t)$. Since both signals are at rest before
contact, zero initial conditions are used.

Two estimation strategies are compared. In the \emph{ensemble-mean} (EM)
strategy, the impacts are averaged, $\bar u=\frac1{N_I}\sum_i u_i$,
$\bar y=\frac1{N_I}\sum_i y_i$, and $\theta$ is the least-squares
estimate on the single averaged transient ($n_t=56$ samples). This is
the approach of \citet{Martin2023}, where the model is estimated with
\texttt{arx} (System Identification Toolbox of Matlab) with stability enforcement. In the \emph{multi-experiment} (ME) strategy, the
regressions of all impacts are stacked,
\begin{equation}
  \hat\theta=\arg\min_\theta\sum_{i=1}^{N_I}\sum_{t=1}^{n_t}
  \bigl(y_i(t)-\varphi_i(t)^\top\theta\bigr)^2,
\end{equation}
which, for a linear model, uses the same information as averaging when
the inputs are identical, but exploits the impact-to-impact variations
of the input and multiplies the number of equations by $N_I$. In both
cases $n_a=n_b=n$ and the delay $n_k\in\{1,\dots,20\}$ is chosen to
minimise the estimation error $\sum_t(y(t)-\hat y_s(t))^2$ on the data,
where $\hat y_s=q^{-n_k}\hat B/\hat A\,u$ is the simulated output (on
$\bar u$ for EM, summed over the impacts for ME), $\hat A(q^{-1})$ and $\hat B(q^{-1})$ being the estimates parameterized by $\hat \theta$.

With a dozen impacts per condition, a separate
validation trial is not available. We use leave-one-impact-out (LOIO)
cross-validation: for each $i$, the model (delay and parameters, for a
given order) is estimated without impact $i$ and simulated on $u_i$, and
\begin{equation}
  \mathrm{FIT}_i = 100\Bigl(1-\frac{\|y_i-\hat y_{s,i}\|_2}{\|y_i-\bar y_i\mathbf 1\|_2}\Bigr)
  \ [\%]
\end{equation}
is computed, $\bar y_i$ being the mean of $y_i$. The median of
$\mathrm{FIT}_i$ over $i$ summarises a data set. Because the body does
not respond identically to two similar pushes, even a perfect model
cannot reach 100\%. To calibrate the fits, we compute a
\emph{repeatability bound}: the fit obtained when $y_i$ is predicted by
the mean response of the other impacts, $\frac{1}{N_I-1}\sum_{j\neq
i}y_j$, a predictor that ignores the input. A model that approaches this
bound captures most of the reproducible part of the response. The order
selection by LOIO is compared with Akaike's (AIC) and the Bayesian (BIC)
information criteria computed on the stacked prediction errors
\citep{Akaike1974,Ljung1999}. Finally, the length of the post-contact
window, 0.5\,s in the earlier analyses, is varied from 0.25 to 1\,s to
assess its influence on the accuracy and robustness of both strategies.

To test whether muscle activity explains the
part of the force response that the transferred work does not, the ARX
model \eqref{eq:arx} is extended with the phasic EMG envelopes $e_c$ of
a set $\mathcal C$ of muscles (arms, legs, or both),
\begin{multline}
  A(q^{-1})\,y(t) = q^{-n_k}B(q^{-1})\,u(t)\\
  + \sum_{c\in\mathcal C} q^{-n_e}B_c(q^{-1})\,e_c(t) + e(t),
  \label{eq:miso}
\end{multline}
with $n_a=n_b=10$, $n_k$ taken from the single-input model, four
coefficients per $B_c$ and a common EMG delay $n_e\in\{0,\dots,10\}$
selected like $n_k$. Since the EMG is itself a response to the push,
\eqref{eq:miso} is not a causal description of the neuromuscular loop
but a test of the predictive information carried by the EMG. It is
estimated with the ME strategy and validated by LOIO.

Finally, the models are compared through model-based descriptors. For
$n\geq 4$ classical analysis based on the individual poles and
zeros have no direct physical meaning and are sensitive to small data changes. We therefore
characterise each model by its response $\hat y_{\mathrm{ref}}$ to a
common reference input $u_{\mathrm{ref}}$, the average over all data
sets of the mean work profile normalised to a final value of 1\,J. From
$\hat y_{\mathrm{ref}}$ we extract the \emph{force gain}
$g=\max_t\hat y_{\mathrm{ref}}(t)$ (N/J), i.e.\ the peak antero-posterior force change per
joule transferred, and the \emph{time to peak} $t_{\mathrm{pk}}$. The
third descriptor is the delay $\tau=n_kT_s$, given by the
optimisation.

\section{Results}\label{sec:results}

\subsection{Data sets and model selection}

In the raw files, the two feet-together folders of S5 contain the same
recording, identified by its name as the eyes-open trial. CFCENL of S5 is
therefore missing, which leaves 39 data sets with 10 to 16 impacts each
(482 impacts). The first pendulum marker was valid in 32 of these 39
trials. Otherwise, the second, third or fifth marker was used.

\begin{figure}
  \centering
  \includegraphics[width=\columnwidth]{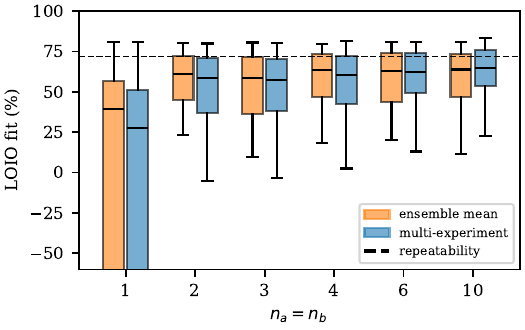}
  \caption{Leave-one-impact-out fit (median per data set, box over the
  39 data sets) versus model order, for estimation on the ensemble mean
  and multi-experiment estimation. Dashed: median repeatability bound.}
  \label{fig:cv}
\end{figure}

Fig.~\ref{fig:cv} shows the LOIO fits as a function of $n$ for both
strategies. First-order models fail. With ME the fit increases
steadily with the order, from 58\% for $n=2$ to 65\% for $n=10$, the
order used by \citet{Martin2023} (median over the 39 data sets). The
median repeatability bound is 72\% (inter-quartile range 62--78\%). This
bound is the fit obtained by predicting each impact with the mean
response of the others: it measures how reproducible the response is,
and no input-driven model can be expected to exceed it by much. The
ratio between the LOIO fit of a data set and its bound has a median of
0.93. The models thus reproduce the response of an unseen impact almost
as well as the average of all other responses, which is the best
prediction available without modelling the impact-to-impact
variability. With EM
the median is similar (64\% for $n=10$), but the two strategies differ
in robustness: with $n=10$, the unconstrained least-squares estimate on
the averaged transient is unstable for 7 of the 39 data sets, which is
why stability had to be enforced in \citet{Martin2023}, and its median
LOIO fit is negative for 3 data sets. None of the ME estimates is
unstable and none has a negative fit, since 10 to 16 transients provide
560 to 900 equations for 20 parameters instead of 56. For comparison,
with the norm-based output of the earlier analyses, the median fit of
the ME models is 49\%, for a repeatability bound of 60\%.

Information criteria are of little help here. With ME, AIC selects
$n\geq 10$ for 33 of the 39 data sets, and the largest order tested
($n=12$) for 21 of them. BIC is more parsimonious (median $n=5$,
inter-quartile range 3--8). Both rely on white prediction errors and
independent samples, whereas the equation errors of consecutive
samples, and of the impacts of a trial, are strongly correlated. LOIO
does not rely on these assumptions. We retain ME with $n=10$, at the top
of Fig.~\ref{fig:cv}, and check that the conclusions below hold with
$n=4$ and $n=6$.

The model accuracy is mainly limited by the repeatability of the
response: across data sets, the LOIO fit is strongly correlated with the
repeatability bound (Spearman $\rho=0.80$, $p<10^{-8}$), and only weakly
with the coefficient of variation of the transferred work ($\rho=-0.26$,
$p=0.11$). Replacing $u_i$ by the energy-consistent input $\Delta E_i$
of \eqref{eq:decomp} lowers the median LOIO fit from 65\% to 61\% (lower
in 27 of 39 data sets, $p=0.003$): the roughness term, although not an
energy, carries some information on the abruptness of the deceleration.

The length of the analysis window matters for both accuracy and
robustness (Table~\ref{tab:window}). With 0.25\,s after contact, the
window ends before the force peak. The response is then only partly
observed, the EM estimate is unstable for 37 of the 39 data sets, and
the ME fit falls to 81\% of the repeatability bound. Between 0.35 and
0.5\,s, the ME models reach 92--93\% of the bound. Longer windows add
the slow return of the force, which is less reproducible (lower bound)
and less well explained by the transferred work (ratio 0.90), but they
provide enough samples for the EM estimate to become stable. 
The effect of vision on the delay is found for all window lengths: the delay is longer with eyes closed in 11 to 14 of the 19 subject-level OE/CE pairs, and shorter in at most 2 (Wilcoxon $p\leq 0.006$), although the median increase varies with the window (20 to 70\,ms).
The
effect of the posture on the gain (Section~\ref{sec:comp}) holds for windows from
0.35 to 0.75\,s ($p\leq 0.037$), but not with 0.25\,s, where the peak
is not observed, nor with 1\,s ($p=0.63$), where the slow postural
corrections dominate the response. The window of 0.5\,s thus appears as
a good compromise, and is kept.

\begin{table}
\caption{Influence of the post-contact window length: median LOIO fit
of the ME models ($n=10$), median repeatability bound, median ratio of
the two, and number of unstable EM estimates (out of 39).}
\label{tab:window}
\centering\footnotesize\setlength{\tabcolsep}{3pt}
\begin{tabular}{cccccc}
\toprule
Window (s) & $n_t$ & ME fit (\%) & Bound (\%) & Ratio & EM unstable\\
\midrule
0.25 & 31  & 53 & 64 & 0.81 & 37\\
0.35 & 41  & 66 & 74 & 0.92 & 16\\
0.50 & 56  & 65 & 72 & 0.93 & 7\\
0.75 & 81  & 61 & 67 & 0.90 & 0\\
1.00 & 106 & 61 & 67 & 0.90 & 0\\
\bottomrule
\end{tabular}
\end{table}

\begin{figure*}
  \centering
  \includegraphics[width=0.95\textwidth]{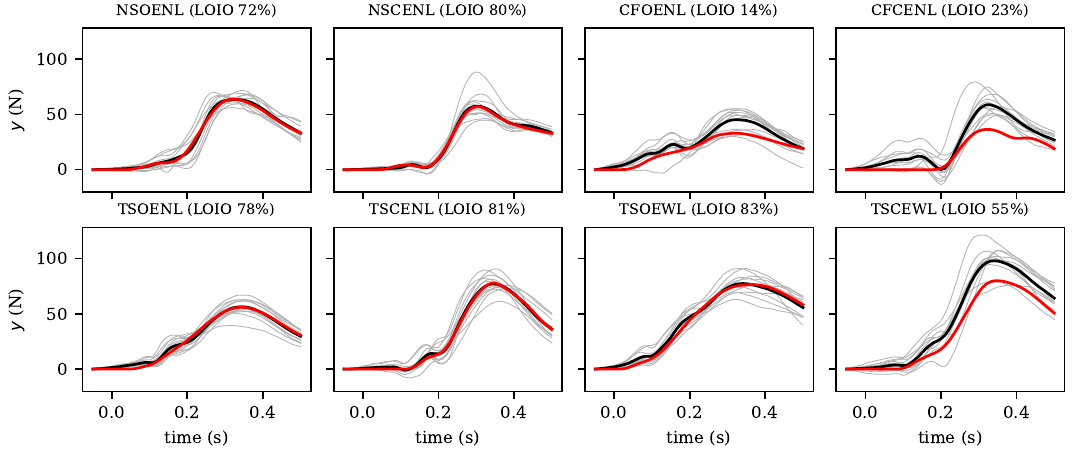}
  \caption{Subject S1: individual antero-posterior GRF responses (grey),
  their mean (black) and the ME model ($n=10$) simulated on the mean
  input (red). The LOIO fit is given in each title.}
  \label{fig:s1}
\end{figure*}

Fig.~\ref{fig:s1} illustrates the models of S1 in the eight conditions.
The antero-posterior force starts to rise about 0.1\,s after contact,
peaks 0.3 to 0.4\,s after contact and decays slowly. This shape is
reproduced with LOIO fits of 55 to 83\% in the feet-apart conditions.
The two feet-together conditions are the least repeatable and the least
well modelled (14 and 23\%). The narrow base of support makes the stance
unstable, and the body then tends to sway like an inverted pendulum
after the push. This makes the response more variable and partly
nonlinear.

\subsection{Comparison of the conditions}\label{sec:comp}

\begin{figure}
  \centering
  \includegraphics[width=\columnwidth]{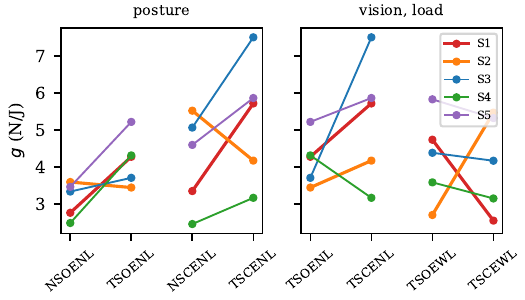}
  \caption{Force gain $g$ of the ME models ($n=10$) for each subject.
  Left: normal stance versus Tai Chi posture, eyes open and closed.
  Right: effect of vision and load in the Tai Chi posture. Thick lines:
  experts (S1, S2).}
  \label{fig:gain}
\end{figure}

Table~\ref{tab:desc} summarises the descriptors of the ME models over
the subjects, and Fig.~\ref{fig:gain} shows the force gain of each
subject. The conditions are compared with two rank-based tests suited to
small samples without normality assumption: the Wilcoxon signed-rank
test for paired differences \citep{Wilcoxon1945} and the Friedman test
for more than two related conditions \citep{Friedman1937}.

Closing the eyes lengthens the identified delay: from 10
to 30\,ms on average with eyes open to 60 to 95\,ms with eyes closed.
Over the 19 subject-level pairs, the delay increases in 14 cases and
never decreases (median $+70$\,ms, Wilcoxon $p=0.001$, although the
pairs of a same subject are not independent). The mean increase is
positive for all five subjects, and the result holds with $n=4$ and
$n=6$. The force gain also increases with eyes closed (13 of 19 pairs,
median $+0.65$\,N/J, $p=0.036$, positive for 4 of the 5 subjects).
Without vision, the response is later and larger.

The Tai Chi posture increases the force gain with
respect to normal stance with feet apart: $g$ is larger in 8 of the 10
(subject, vision) pairs (median $+1.4$\,N/J, $p=0.027$), for four of the
five subjects, and a Friedman test over the eight conditions (S1--S4,
complete data) detects a condition effect ($p=0.038$). Both results
hold with $n=4$ and $n=6$. Standing with the feet together tends to
lower the gain (7 of 9 pairs, $p=0.07$). The heavier pendulum does not
change the gain per joule (4 of 10 pairs higher, $p=0.43$), which
supports the linear scaling of the response with the transferred
energy. The delay and the time to peak do not depend on the posture.

The gain is most
reproducible across subjects in the two unloaded eyes-open conditions
with feet apart (coefficient of variation 15\% in NSOENL, 16\% in
TSOENL), against 28 to 48\% in all the others. With only two experts, no
difference between groups can be established: the only subject whose
gain does not increase in the Tai Chi posture is one of the experts
(S2), and no descriptor separates the groups.

\begin{table*}
\caption{Descriptors of the multi-experiment models ($n=10$), mean
$\pm$ standard deviation over subjects: number of subjects $N_S$, number
of impacts $N_I$, delay $\tau$, median LOIO fit, force gain $g$ and
time to peak $t_{\mathrm{pk}}$ of the response to the reference input.}
\label{tab:desc}
\centering
\begin{tabular}{lcccccc}
\toprule
Condition & $N_S$ & $N_I$ & $\tau$ (ms) & LOIO fit (\%) & $g$ (N/J) & $t_{\mathrm{pk}}$ (ms)\\
\midrule
NSOENL  & 5 & 11--16 & 32 $\pm$ 29 & 64 $\pm$ 22 & 3.1 $\pm$ 0.5 & 398 $\pm$ 58 \\
NSCENL  & 5 & 11--15 & 64 $\pm$ 38 & 74 $\pm$ 9 & 4.2 $\pm$ 1.3 & 392 $\pm$ 79 \\
CFOENL  & 5 & 10--14 & 10 $\pm$ 0 & 23 $\pm$ 19 & 2.5 $\pm$ 1.0 & 382 $\pm$ 48 \\
CFCENL  & 4 & 11--15 & 95 $\pm$ 66 & 55 $\pm$ 21 & 3.1 $\pm$ 1.5 & 380 $\pm$ 29 \\
TSOENL  & 5 & 12--15 & 22 $\pm$ 16 & 60 $\pm$ 20 & 4.2 $\pm$ 0.7 & 356 $\pm$ 38 \\
TSCENL  & 5 & 10--15 & 58 $\pm$ 54 & 59 $\pm$ 24 & 5.3 $\pm$ 1.7 & 386 $\pm$ 26 \\
TSOEWL  & 5 & 10--13 & 10 $\pm$ 0 & 64 $\pm$ 19 & 4.2 $\pm$ 1.2 & 376 $\pm$ 43 \\
TSCEWL  & 5 & 10--13 & 60 $\pm$ 28 & 65 $\pm$ 8 & 4.1 $\pm$ 1.3 & 376 $\pm$ 48 \\
\bottomrule
\end{tabular}
\end{table*}

\subsection{Muscle activity}

\begin{figure*}
  \centering
  \includegraphics[width=0.95\textwidth]{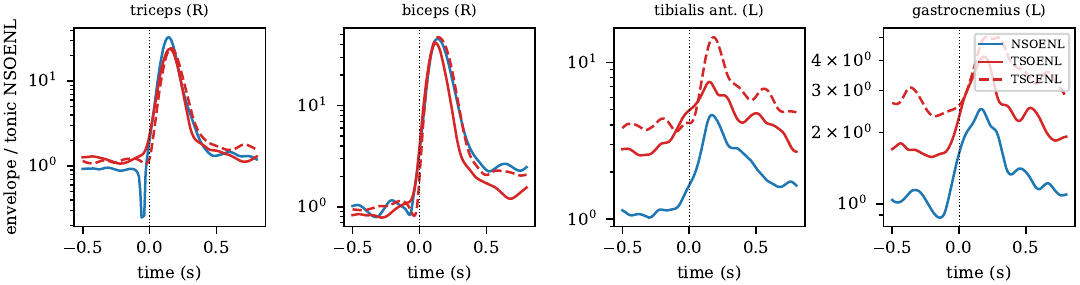}
  \caption{EMG envelopes around the impacts (contact at $t=0$), averaged
  over impacts and subjects (geometric mean) and normalised by the tonic
  activity of each subject in NSOENL: normal stance, eyes open (blue);
  Tai Chi posture, eyes open (red) and closed (red, dashed).}
  \label{fig:emg}
\end{figure*}

Fig.~\ref{fig:emg} shows the EMG envelopes averaged over impacts and
subjects (geometric mean), normalised by the tonic activity of each
subject in NSOENL.

The Tai Chi posture raises the tonic activity of
both lower-leg muscles with respect to normal stance: by a median factor
of 1.75 for the tibialis anterior (higher in 8 of the 10 subject--vision
pairs, Wilcoxon $p=0.02$) and 1.48 for the gastrocnemius (9 of 10,
$p=0.006$), whereas the arm muscles change little (factors 1.10 and 1.08,
$p>0.15$). Standing with the feet together has a similar but weaker
effect (1.5 and 1.2 with eyes open). The flexed posture is thus held with
a co-activation of the ankle agonist and antagonist, i.e.\ with an
increased ankle stiffness.

Each impact produces a large burst in the arm
muscles, peaking 110--160\,ms after contact (15 to 100 times the
pre-impact level), followed by smaller leg responses (2 to 7 times)
peaking 150--300\,ms after contact, i.e.\ before the force peak. At
100\,Hz, the arm burst cannot be fully separated from a movement
artefact of the electrodes under the impact, although its peak is
clearly later than the contact. In the Tai Chi posture the triceps burst
is smaller than in normal stance (median factor 0.62, 9 of 10 pairs,
$p=0.014$), and the tibialis burst, relative to its higher tonic level,
tends to be smaller as well (0.69, $p=0.08$), so that its absolute peak
is about the same. Closing the eyes tends to increase the tibialis burst
(13 of 19 pairs, median factor 1.6, $p=0.055$). No EMG feature separates
the two experts from the other subjects.

To test whether the responses change
during a trial, the Spearman correlation between the impact rank and the
peak of each signal is computed per data set. Neither the force peak
(median $\rho=-0.07$, negative in 22 of 39 data sets, $p=0.51$) nor the
energy lost by the pendulum ($\rho=-0.03$) drifts, and the arm bursts do
not decrease. The tibialis anterior burst, in contrast, decreases in 30
of the 39 trials (median $\rho=-0.35$, Wilcoxon $p=2\cdot10^{-5}$), by
about 46\% from the first to the last impact according to a linear fit.
The decrease is concentrated on the first impacts (the first burst is
1.5 times the median of the following ones, and without the first two
impacts the trend is only marginal, $p=0.09$), is present in all
postures and in both groups, and is stronger with eyes closed. It is
thus an initial over-reaction, damped once the perturbation becomes
predictable, which does not affect the force transmitted to the ground.

Adding the phasic envelopes to the model
\eqref{eq:miso} does not improve the LOIO fit: the median change is
$-1.4$ points with the leg muscles (improvement in 15 of 39 data sets,
$p=0.34$) and $0.0$ with the arm muscles (19 of 39, $p=0.24$). Large
gains occur only in a few trials containing an atypical impact (e.g.\
S2, TSCENL). Once the force output is well defined, the transferred work
thus explains the reproducible part of the response, and the EMG
envelopes recorded at 100\,Hz bring no additional predictive
information.

\section{Discussion}\label{sec:discussion}

The main improvement over the earlier analyses of these data
\citep{NYo2019,Martin2023} concerns the definition of the signals.
Relating the discrete work to the energy lost by the pendulum
\eqref{eq:decomp} explains most of the input variability. Defining the
output component-wise makes it independent of the zeroing of the force
plate. This second change has the largest effect: it raises the median
validation fit from 49\% to 65\% and gives smooth responses whose shape
is shared by all subjects. More generally, when a force plate is zeroed
with the subject on it, nonlinear functions of the force, such as the
norm, should be computed on absolute forces or replaced by
component-wise outputs.

Averaging the
transients before identification is appealing because it reduces the
noise, but for a model with 20 parameters it leaves only 56 equations
and discards the input variability between impacts. The resulting
estimates can be unstable and are fragile under cross-validation.
Stacking the impacts removes both problems at no computational cost. In
both cases the ARX estimate is biased if the equation error is coloured
\citep{Ljung1999,SoderstromStoica1989}. Output-error or
instrumental-variable estimators, and an errors-in-variables treatment
of the input computed from marker data, are natural next steps. A
smoothed estimate of the deceleration profile, or the contact force
measured by a load cell on the pendulum, would also give a cleaner
input.

Taken together, the models and the EMG
describe a consistent strategy in the Tai Chi posture. The ankle is
stiffened in advance by co-activation, the reactive arm burst is
smaller, and more of the push is transmitted to the ground per joule
transferred. The body thus behaves as a stiffer, better-coupled
structure that takes the push through the support, rather than yielding
to it. With eyes closed, the response is both later and larger. The
fast habituation of the tibialis burst, with an unchanged force output,
shows that part of the reactive leg activity is an over-reaction to the
first, unfamiliar pushes, and supports the use of all impacts in the
estimation. With five subjects and two experts, these results remain
exploratory. A definitive analysis will require a larger cohort, EMG
sampled at 1--2\,kHz and normalised to maximal voluntary contractions,
and the thigh and trunk muscles that hold the flexed posture.

The pendulum impacts excite a narrow band and
vary little in amplitude, which limits the identifiability of
higher-order and nonlinear models. Recording and varying the release
height would both enrich the excitation and allow testing the linearity
of the force response. Finally, the stance is under active
neuromuscular control, so that separating passive mechanics from
feedback would require a closed-loop identification set-up
\citep{vanderKooij2005}.

\section{Conclusion}\label{sec:conclusion}

We have identified linear models from the work transferred by a pendulum
hitting the forearm to the antero-posterior ground reaction force, for
five subjects in eight postural conditions, using repeated transients
of 0.55\,s. Relating the input to the energy of the pendulum and
defining the output component-wise proved essential. Stacking
the impacts in a multi-experiment least-squares problem yields stable
tenth-order ARX models whose leave-one-impact-out fit reaches 93\% of
the repeatability of the response itself. The models show a longer
delay and a larger gain with eyes closed, and a larger force gain in the
Tai Chi posture, which the EMG relates to an anticipatory co-activation
of the ankle muscles and a smaller reactive arm burst. Future work will
address bias-free estimators, a better-conditioned input, the vertical
force as a second output, EMG sampled at an adequate rate, and a larger
cohort with contrasted levels of expertise, analysed with mixed-effects
models.

\begin{ack}
The authors thank the volunteers, the staff of the GIPSA-lab
biomechanics platform, and Sufiyan N'Yo for his contribution to the
experimental campaign and to the preliminary analysis during his
Master's thesis. The authors used Claude (Anthropic) to assist with the
data analysis, the Python implementation and the writing; they verified
all results and take full responsibility for the content.
\end{ack}

\bibliography{taichi_sysid}

\end{document}